\documentclass[reprint,aps,prl,twocolumn,superscriptaddress,showpacs,floatfix]{revtex4-1}

\usepackage{graphicx,enumerate}
\usepackage{amsmath,epsfig}

\newcommand{\be}{\begin{equation}}
\newcommand{\ee}{\end{equation}}
\newcommand{\ba}{\begin{eqnarray}}
\newcommand{\ea}{\end{eqnarray}}

\begin{document}

\title{Testing general relativity with accretion-flow imaging of Sgr~A$^\ast$}

\author{Tim Johannsen}
\affiliation{Perimeter Institute for Theoretical Physics, Waterloo, Ontario N2L 2Y5, Canada}
\affiliation{Department of Physics and Astronomy, University of Waterloo, Waterloo, Ontario N2L 3G1, Canada}
\affiliation{Canadian Institute for Theoretical Astrophysics, University of Toronto, Toronto, Ontario M5S 3H8, Canada}

\author{Carlos Wang}
\affiliation{Department of Physics and Astronomy, University of Waterloo, Waterloo, Ontario N2L 3G1, Canada}

\author{Avery E. Broderick}
\affiliation{Perimeter Institute for Theoretical Physics, Waterloo, Ontario N2L 2Y5, Canada}
\affiliation{Department of Physics and Astronomy, University of Waterloo, Waterloo, Ontario N2L 3G1, Canada}

\author{Sheperd~S.~Doeleman}
\affiliation{MIT Haystack Observatory, Westford, Massachusetts 01886, USA}
\affiliation{Harvard-Smithsonian Center for Astrophysics, 60 Garden Street, Cambridge, Massachusetts 02138, USA}

\author{Vincent L. Fish}
\affiliation{MIT Haystack Observatory, Westford, Massachusetts 01886, USA}

\author{Abraham Loeb}
\affiliation{Department of Astronomy, Harvard University, 60 Garden Street, Cambridge, Massachusetts 02138, USA}

\author{Dimitrios Psaltis}
\affiliation{Astronomy Department, University of Arizona, 933 North Cherry Avenue, Tucson, Arizona 85721, USA}

\begin{abstract}

The Event Horizon Telescope is a global very-long baseline interferometer capable of probing potential deviations from the Kerr metric, which is believed to provide the unique description of astrophysical black holes. Here we report an updated constraint on the quadrupolar deviation of Sagittarius~A$^\ast$ within the context of a radiatively inefficient accretion flow model in a quasi-Kerr background. We also simulate near-future constraints obtainable by the forthcoming eight-station array and show that in this model already a one-day observation can measure the spin magnitude to within $0.005$, the inclination to within $0.09^\circ$, the position angle to within $0.04^\circ$, and the quadrupolar deviation to within $0.005$ at $3\sigma$ confidence. Thus, we are entering an era of high-precision strong gravity measurements.

\end{abstract}


\maketitle


The supermassive black holes Sgr~A$^\ast$ and in M87 are the prime targets of the Event Horizon Telescope (EHT). These sources have already been observed with a three-station array, comprised by the James Clerk Maxwell Telescope and the Sub-Millimeter Array in Hawaii (Hawaii), the Submillimeter Telescope Observatory in Arizona (SMT), and several dishes of the Combined Array for Research in Millimeter-wave Astronomy (CARMA) in California, which has resolved structures on scales of only $4r_S$ in Sagittarius~A$^\ast$ (Sgr~A$^\ast$)~\cite{Doele08} and $5.5r_S$ in M87~\cite{Doele12}, respectively. Here, $r_S\equiv 2r_g\equiv 2GM/c^2$ is the Schwarzschild radius of a black hole with mass $M$, and $r_g$ is its gravitational radius~\cite{footnote}.

According to the general-relativistic no-hair theorem, stationary, electrically neutral black holes in vacuum only depend on their masses $M$ and spins $J$ and are uniquely described by the Kerr metric~\cite{nocharge}. Mass and spin are the first two multipole moments of the Kerr metric, and all higher-order moments can be expressed in terms of them by the relation $M_l + iS_l = M(ia)^l$, where $M_l$ and $S_l$ are the mass and current multipole moments, respectively, and $a\equiv J/M$ is the spin parameter (see, e.g., Ref.~\cite{heu96}).

General relativity has been well confirmed in the regime of weak gravitational fields~\cite{will06}, but still remains practically untested in the strong-field regime found around compact objects~\cite{psalLRR}. It is possible to test the no-hair theorem using parametrically deformed Kerr-like spacetimes that depend on one or more free parameters in addition to mass and spin. Observations may then be used to measure the deviations. If none are detected, the compact object is consistent with a Kerr black hole. If, however, nonzero deviations are measured, there are two possible interpretations. If general relativity still holds, the object is not a black hole but, instead, another stable stellar configuration or some exotic object. Otherwise, the no-hair theorem would be falsified. 

By design, parametric Kerr-like spacetimes encompass many theories of gravity at once and generally do not derive from the action of any particular theory. It is assumed, however, that particles move along geodesics. Tests of the no-hair theorem in a Kerr-like spacetime have been suggested for gravitational-wave observations of extreme mass-ratio inspirals~\cite{gair13} and electromagnetic observations of accretion flows surrounding black holes (e.g., \cite{PaperII,EM}). Other tests of the no-hair theorem include electromagnetic observations of pulsar black hole binaries~\cite{pulsars,PWK16} and stars on orbits around Sgr~A$^\ast$~\cite{Sgr,PWK16}, though these constitute weak-field probes.

Here we employ a quasi-Kerr metric~\cite{GB06}, which modifies the quadrupole moment $Q_{\rm K}\equiv M_2=-Ma^2$ of the Kerr metric according to the equation $Q_{\rm QK} = -M(a^2 + \epsilon M^2)$, where $\epsilon$ is a dimensionless parameter that measures potential deviations from the Kerr metric. The quasi-Kerr metric is of the form $g_{\rm \mu\nu}^{\rm QK} = g_{\rm \mu\nu}^{\rm K} + \epsilon h_{\mu\nu}$, where $g_{\mu\nu}^{\rm K}$ is the Kerr metric and $h_{\mu\nu}$ is diagonal. An explicit expression of this metric is given in Ref.~\cite{GB06}. Note that the expression for the quadrupole moment is exact for sufficiently small values of the spin and the parameter $\epsilon$, but it may only be approximate otherwise~\cite{supp}.

A key objective of the EHT is to produce the first direct image of a black hole. These typically reveal a dark region at the center, the so-called shadow~\cite{shadow}. The shape of this shadow is exactly circular for a Schwarzschild black hole and nearly circular for a Kerr black hole unless its spin is very large and the inclination is high. However, the shape of the shadow becomes asymmetric if the no-hair theorem is violated, e.g., for nonzero values of the parameter $\epsilon$ in the quasi-Kerr metric~\cite{PaperII}.

Sgr~A$^\ast$ is the black hole with the largest angular cross section in the sky. While several models for its accretion flow exist~\cite{flowmodels}, these typically fall within the radiatively inefficient accretion flow paradigm (RIAF). A recent analysis within the context of RIAFs found that images of accretion flows in the quasi-Kerr spacetime differ significantly from those in a Kerr background and, already, these may be grossly distinguished by early EHT data. Furthermore, measurements of the inclination and spin position angle are robust to the inclusion of a quadrupolar deviation from the Kerr metric~\cite{Bro14}. In particular, Ref.~\cite{Bro14} obtained the $1\sigma$ constraints on the spin magnitude \mbox{$a_\ast=0^{+0.7}_{-0.0}$}, inclination \mbox{$\theta=65^{\circ+21^\circ}_{~-11^\circ}$}, and orientation \mbox{$\xi=127^{\circ+17^\circ}_{~-14^\circ}$} (up to a $180^\circ$ degeneracy), while the deviation parameter $\epsilon$ remained unconstrained.


\begin{figure}[ht]
\begin{center}
\psfig{figure=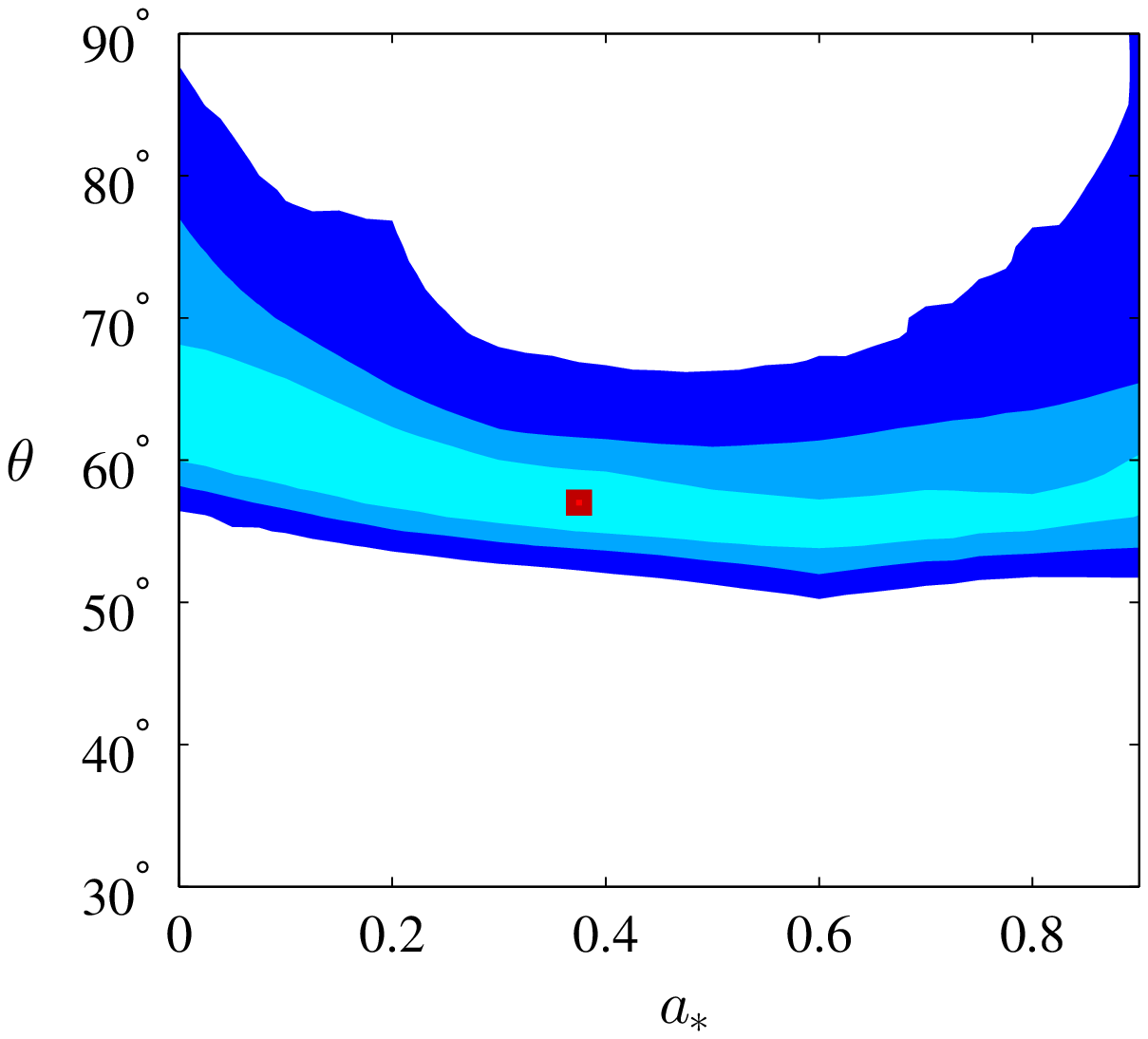,width=0.46\textwidth}
\psfig{figure=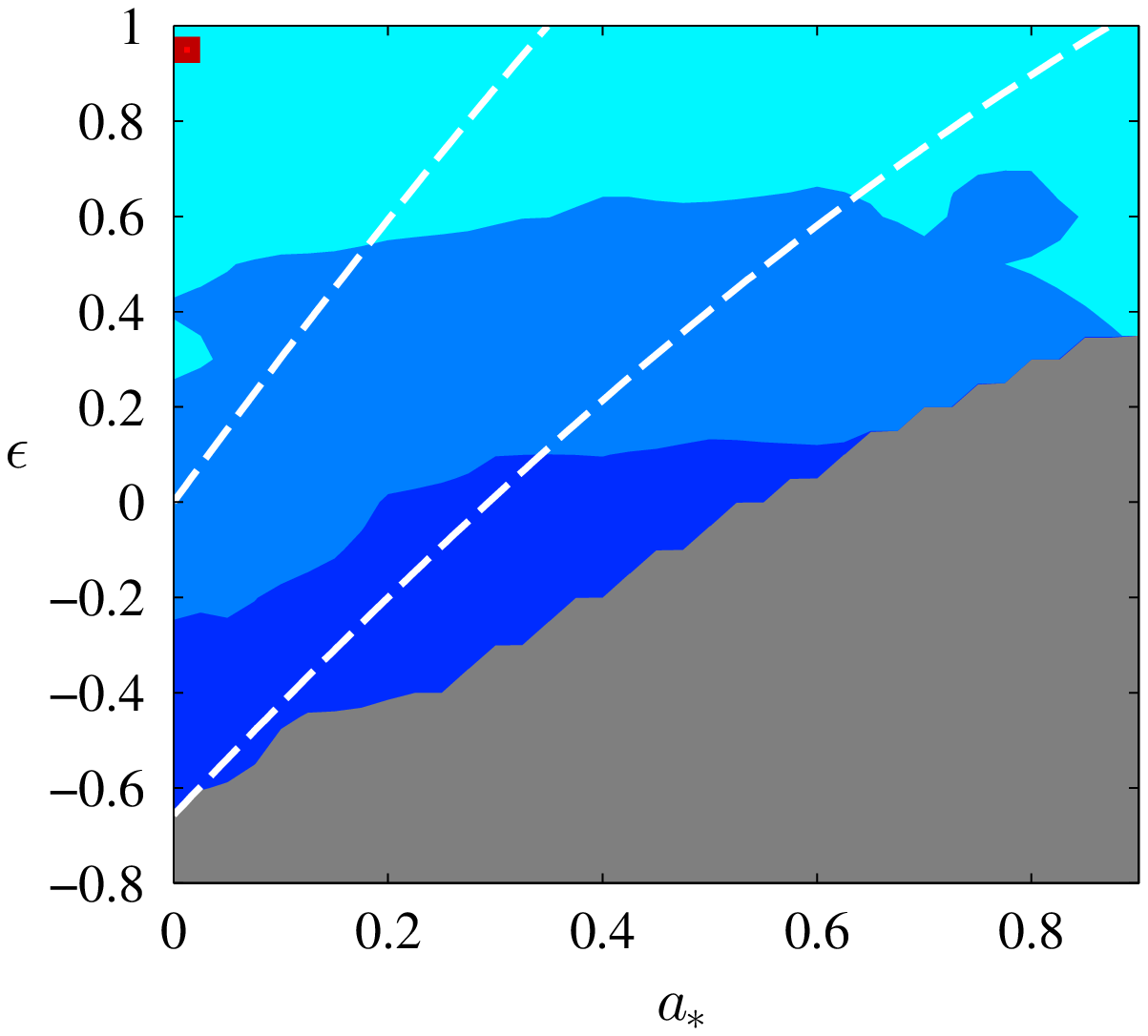,width=0.46\textwidth}
\end{center}
\caption{$1\sigma$, $2\sigma$, and $3\sigma$ confidence contours of the posterior probability density as a function of (top) the spin magnitude $a_\ast$ and inclination $\theta$ and (bottom) the spin magnitude and quadrupolar deviation parameter $\epsilon$, marginalized over all other quantities. The red dot in each panel denotes the maximum of the respective 2D probability density and dashed white lines correspond to constant ISCO radii of (top) $r=6r_g$ and (bottom) $r=5r_g$. The gray region is excluded.}
\label{fig:constraints}
\end{figure}

In April 2009 and in March/April 2011--2013, additional observations of Sgr~A$^\ast$ were carried out at 230~GHz using the same three-station telescope array~\cite{Fish11,Fish16}. A comprehensive analysis of these observations together with updated parameter estimates for Sgr~A$^\ast$ assuming a Kerr background can be found in Ref.~\cite{update}. Here, we focus on the constraints on the quadrupolar deviation parameter. Following the procedure described in Ref.~\cite{Bro14} and allowing for closure phase shifts as in Ref.~\cite{update}, we produced an updated set of parameter estimates within the same parameter space~\cite{supp} using the image library of Ref.~\cite{Bro14} refined by an additional 12,501 images.

Figure~\ref{fig:constraints} shows the spin magnitude--inclination and the spin magnitude--quadrupolar deviation posterior probability distributions, each marginalized over all other parameters (spin orientation angle and, respectively, deviation parameter and inclination). As in Ref.~\cite{Bro14}, the spin magnitude correlates with both the inclination and the deviation parameter, although it is unclear whether the latter correlation is still primarily determined by the location of the innermost stable circular orbit (ISCO).

We obtain new constraints on the spin magnitude \mbox{$a_\ast=0^{+0.90}_{-0.00}$}, inclination \mbox{$\theta=57.0^{\circ+3.0^\circ}_{~-2.0^\circ}$}, orientation \mbox{$\xi=156^{\circ+5^\circ}_{~-4^\circ}$}, and deviation parameter \mbox{$\epsilon=1.00^{+0.00}_{-0.40}$}, where we quote $1\sigma$ errors on the respective posterior probability densities marginalized over all other parameters~\cite{supp,astar}. Formally this implies that Sgr~A$^\ast$ is consistent with a Kerr black hole only at the $3\sigma$ level. However, this constraint on the parameter $\epsilon$ is substantially biased by the restricted range of values of the spin and the quadrupolar deviation we consider affecting the marginalization process and, therefore, overestimating the magnitude of the deviation. A better measure is the 2D probability distribution shown in Fig.~\ref{fig:constraints}, from which it is clear that Sgr~A$^\ast$ is consistent with a Kerr black hole well within the $2\sigma$ level. Even though the maximum of this distribution is located at the edge of the parameter space, the confirmation of the Kerr nature of Sgr~A$^\ast$ at this level should be unaffected by the considered values of the spin and the deviation parameter, because the $1\sigma$ and $2\sigma$ regions are very large. Therefore, we expect that Sgr~A$^\ast$ remains consistent with a Kerr black hole at the $2\sigma$ level even for larger values of the parameter $\epsilon$. However, the quoted confidence intervals should be viewed with caution. Although our result implies that Sgr~A$^\ast$ is in mild tension with being a Kerr black hole, it is most likely dominated by systematic model uncertainties, which do not incorporate other effects such as the vertical structure and variability of the accretion disk, the plasma density and magnetic field strength in and above the disk, and the presence of outflows.

Our constraints on the parameters of Sgr~A$^\ast$ are broadly consistent with the values given in Ref.~\cite{update} and improve upon the constraints on the inclination and spin orientation of Ref.~\cite{Bro14} by roughly a factor of four. In addition, the $180^\circ$ degeneracy of the spin orientation is removed. However, the constraint on the spin magnitude is about 30\% weaker than the constraint of Ref.~\cite{Bro14} and the spin magnitude is now unconstrained. This is in accordance with the results of Refs.~\cite{EHTuncertainties1,EHTuncertainties2} which found that the inclination and spin orientation can be inferred much more precisely from the visibility magnitudes and closure phase data than the spin magnitude.


In 2015, the three-station array Hawaii--SMT--CARMA was expanded to include the Atacama Large Millimeter/submillimeter Array (ALMA) in Chile, the Large Millimeter Telescope in Mexico, the South Pole Telescope (SPT), the Plateau de Bure interferometer in France, and the Pico Veleta Observatory in Spain. Thus, we also assess the prospects of measuring the spin magnitude and position angle, the inclination, and the quadrupolar deviation parameter of Sgr~A$^\ast$ with an eight-station array in the near future. The sensitivity and resolution of this enlarged array will be greatly increased, caused primarily by ALMA which will have a sensitivity that is about 50 times greater than the sensitivity of the current stations and the long baselines from the stations in the northern hemisphere to the SPT. In addition, this array allows for the measurement of closure phases along many different telescope triangles, some of which depend very sensitively on the parameters of Sgr~A$^\ast$~\cite{CP}.

To do this we simulate a single 24 h observing run at 230 GHz using a library image with $a_\ast=0.15$, $\theta=60^\circ$, $\xi=160^\circ$, and $\epsilon=0$, motivated by the results of Ref.~\cite{update}. Simulated visibilities and closure phases are computed every 10 min for all baselines comprised of telescopes for which Sgr A$^\ast$ is above a zenith angle of $70^\circ$. Typical long baseline observing periods are 2-4 h. For each station, we use the system equivalent flux density at this frequency listed in Ref.~\cite{SEFDs}. We assume a 4~GHz recording bandwidth and a 10 s atmospheric correlation time for all measurements. We further assume that the radio emission experiences electron scatter broadening according to the scattering law of Ref.~\cite{Bower}.

\begin{figure}[ht]
\begin{center}
\psfig{figure=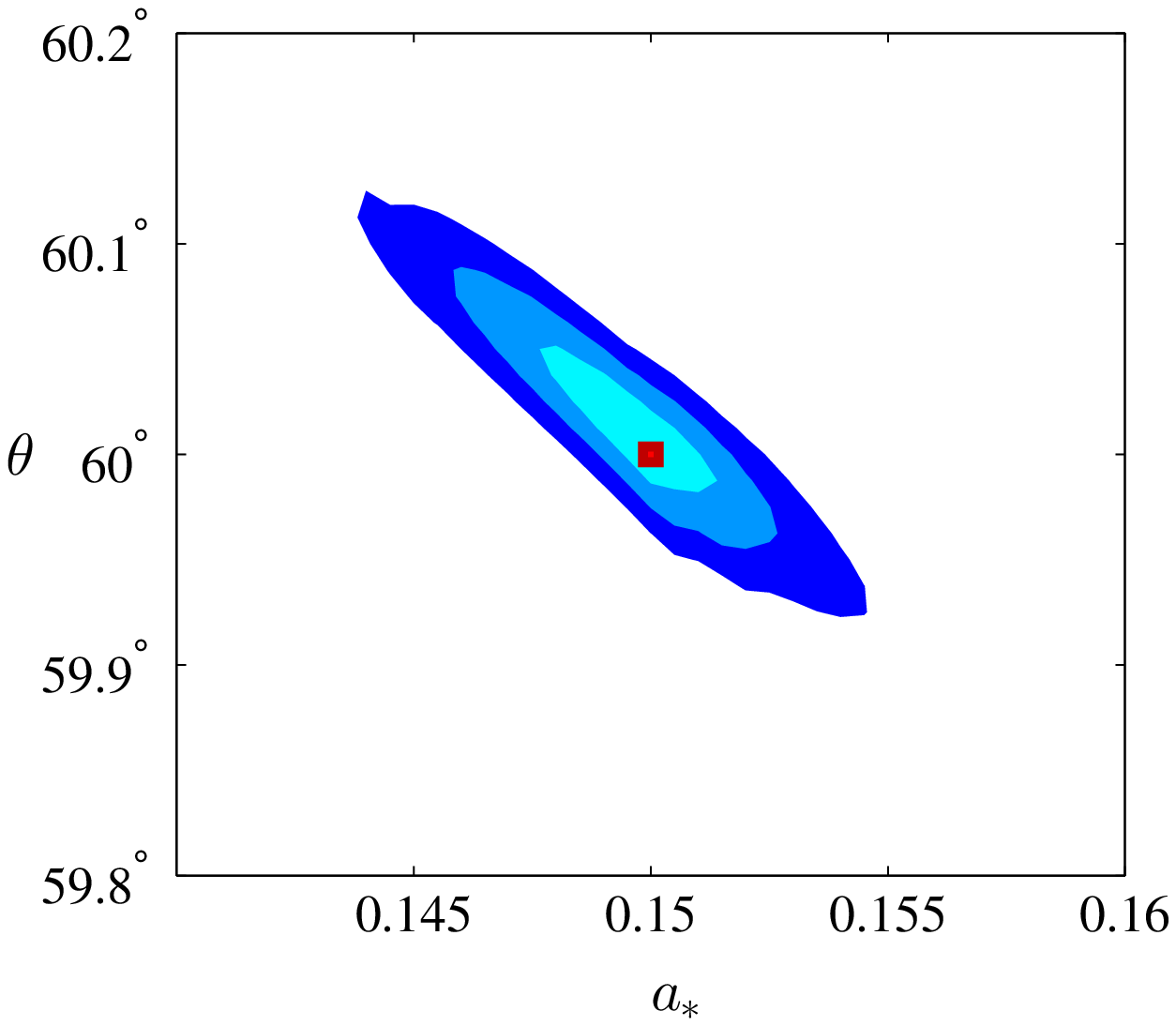,width=0.46\textwidth}
\psfig{figure=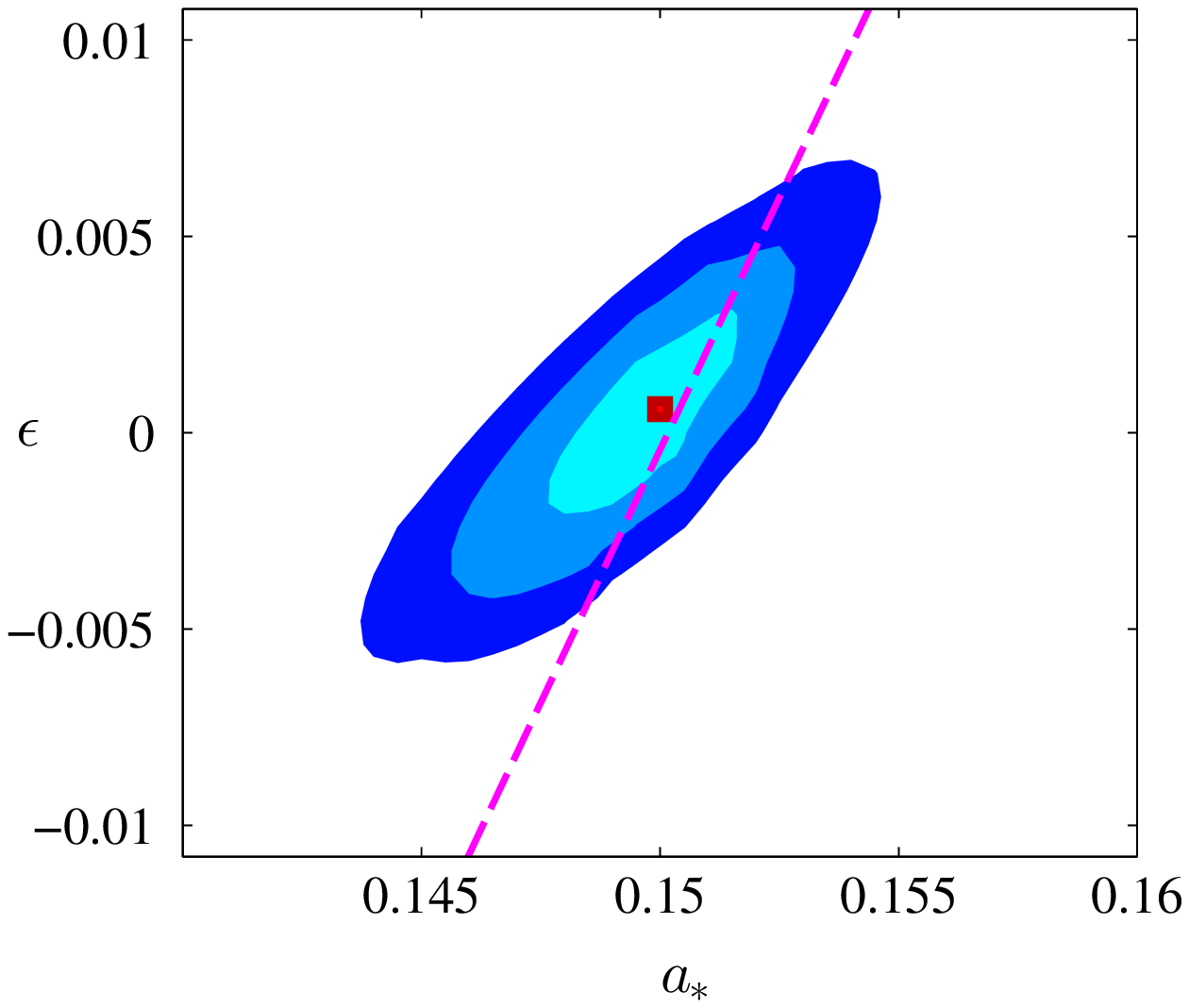,width=0.46\textwidth}
\end{center}
\caption{Simulated posterior probability density as a function of (top) the spin magnitude $a_\ast$ and inclination $\theta$ and (bottom) the spin magnitude and deviation parameter $\epsilon$, marginalized over all other quantities. Solid, dashed, and dotted white lines show the $1\sigma$, $2\sigma$, and $3\sigma$ confidence contours, respectively. The dashed magenta line denotes the location of the ISCO with constant radius $r=5.5r_g$.}
\label{fig:sim}
\end{figure}

For our analysis, we created a new library of RIAF images which consists of a total of 50,061 images with values of the spin magnitude $0.14\leq a_\ast \leq0.16$, inclination $59.8^\circ\leq \theta \leq 60.2^\circ$, and deviation parameter $-0.0108\leq\epsilon\leq0.0108$, varied with the respective step sizes $\Delta a_\ast=0.0005$, $\Delta \theta=0.0125^\circ$, and $\Delta\epsilon=0.0006$. Subsequently, each image was rotated by a set of spin position angles $\xi$ in the range $159.88^\circ\leq\xi\leq160.12^\circ$ in steps of $\Delta\xi=0.005^\circ$. All images were generated using the coefficients of the density and temperature of the thermal and nonthermal distribution of electrons from fits of the radio spectral energy distribution of Sgr~A$^\ast$ obtained in Ref.~\cite{Bro14}. For each library image, an associated likelihood was constructed using the simulated data following the procedure in Ref.~\cite{Bro14}.

Figure~\ref{fig:sim} shows the spin magnitude--inclination and the spin magnitude--quadrupolar deviation posterior probability distributions, respectively marginalized over all other quantities. The spin magnitude remains strongly correlated with the inclination, while the deviation parameter $\epsilon$ is weakly correlated with the inclination. Neither is correlated with the spin position angle.
All parameters in our simulation are tightly constrained: $a_\ast=0.150^{+0.004}_{-0.005}$, $\theta=60.01^{\circ+0.09^\circ}_{~-0.06^\circ}$, $\xi=159.99^\circ\pm0.04^\circ$, and $\epsilon=0\pm0.005$, where we quote $3\sigma$ errors~\cite{supp}. As we show in the bottom panel of Fig.~\ref{fig:sim}, the contours in the $(a_\ast,\epsilon)$ plane are not exactly aligned with lines of constant ISCO radius which suggests that this is not a fundamental degeneracy (cf., the discussion in Ref.~\cite{Review2}).


The reconstructed spacetime parameters are highly precise, despite adopting realistic station performance estimates, and indicate the forthcoming capability of the EHT to probe deviations from general relativity. However, these have been obtained within the context of a specific astrophysical paradigm, placing an as yet poorly understood prior on the analysis which will be investigated elsewhere. The effects of the chosen model and the variability of the accretion flow are difficult to estimate quantitatively at this point, but these will most likely be the dominant source of uncertainty. Such an assessment will require either increasingly parametrized models or further theoretical development of geometrically thick accretion flows in order to account for the astrophysical effects neglected here. Nonetheless, our analysis demonstrates the expected dramatic improvement of the constraints based on observations with a large EHT array given one particular model.

We have also neglected potential systematic errors associated with the uncertain gain calibrations between long and short baselines, estimated in currently reported visibility magnitudes to produce systematic variations of up to 5\%~\cite{Fish11}. However, observations with the larger array will be able to mitigate these through substantially increased sensitivity, redundant baselines, and the construction of closure amplitudes, defined for station quadrangles, which are independent of station-specific gain estimates. Similarly, closure phases are by construction independent of station-specific phase errors, and will benefit from redundant triangles. Likewise, a scheduled increase of recording bandwidth will further increase the array sensitivity by a factor $\sqrt{8}$. At present, there also exist substantial uncertainties in our knowledge of the interstellar scattering law of Ref.~\cite{Bower}, which, however, can be rectified by additional observations and refined modeling~\cite{Psaltis14}. Finally, refractive substructure along the line of sight can cause stochastic variations of the image~\cite{deblurring} which will average out if its blurring effect is sufficiently small~\cite{Lu16} (cf., Ref.~\cite{EHTuncertainties1}). Therefore, we expect the overall impact of these uncertainties on the simulated constraints to be relatively small.

Our results for the spin magnitude and deviation parameter also depend on the mass and distance of Sgr~A$^\ast$ which affect the overall scale of the images and the spectral fits for the electron density and temperature. Current measurements of the mass and distance of Sgr~A$^\ast$ from near-infrared monitoring of stellar orbits close to the black hole have relative errors on the order of 3\% and 1\%, respectively, which are, however, strongly correlated~\cite{Sgrorbits}. These uncertainties will be further reduced by continued monitoring, by the expected improvement in astrometry with the instrument GRAVITY for the Very Large Telescope Interferometer~\cite{GRAVITY}, in combination with EHT measurements of the shadow size of Sgr~A$^\ast$~\cite{Psaltis14,ShadowSize1,ShadowSize2}, and could reach a precision of $\sim$0.1\% with a 30m class telescope~\cite{Weinberg}. Recently, Ref.~\cite{Reid14} independently measured the distance to Sgr~A$^\ast$ to within 2\%.


T.J. was supported by a CITA National Fellowship at the University of Waterloo and is supported in part by Perimeter Institute for Theoretical Physics. A.E.B. receives financial support from Perimeter Institute for Theoretical Physics and the Natural Sciences and Engineering Research Council of Canada through a Discovery Grant. Research at Perimeter Institute is supported by the Government of Canada through Industry Canada and by the Province of Ontario through the Ministry of Research and Innovation. The Event Horizon Telescope is supported by grants from the National Science Foundation and from the Gordon and Betty Moore Foundation (Grant No. GBMF-3561).


\vspace{0.4cm}

\begin{center}
{\large Supplemental Material}
\end{center}
\vspace{0.2cm}

\begin{figure*}[ht]
\begin{center}
\epsfig{figure=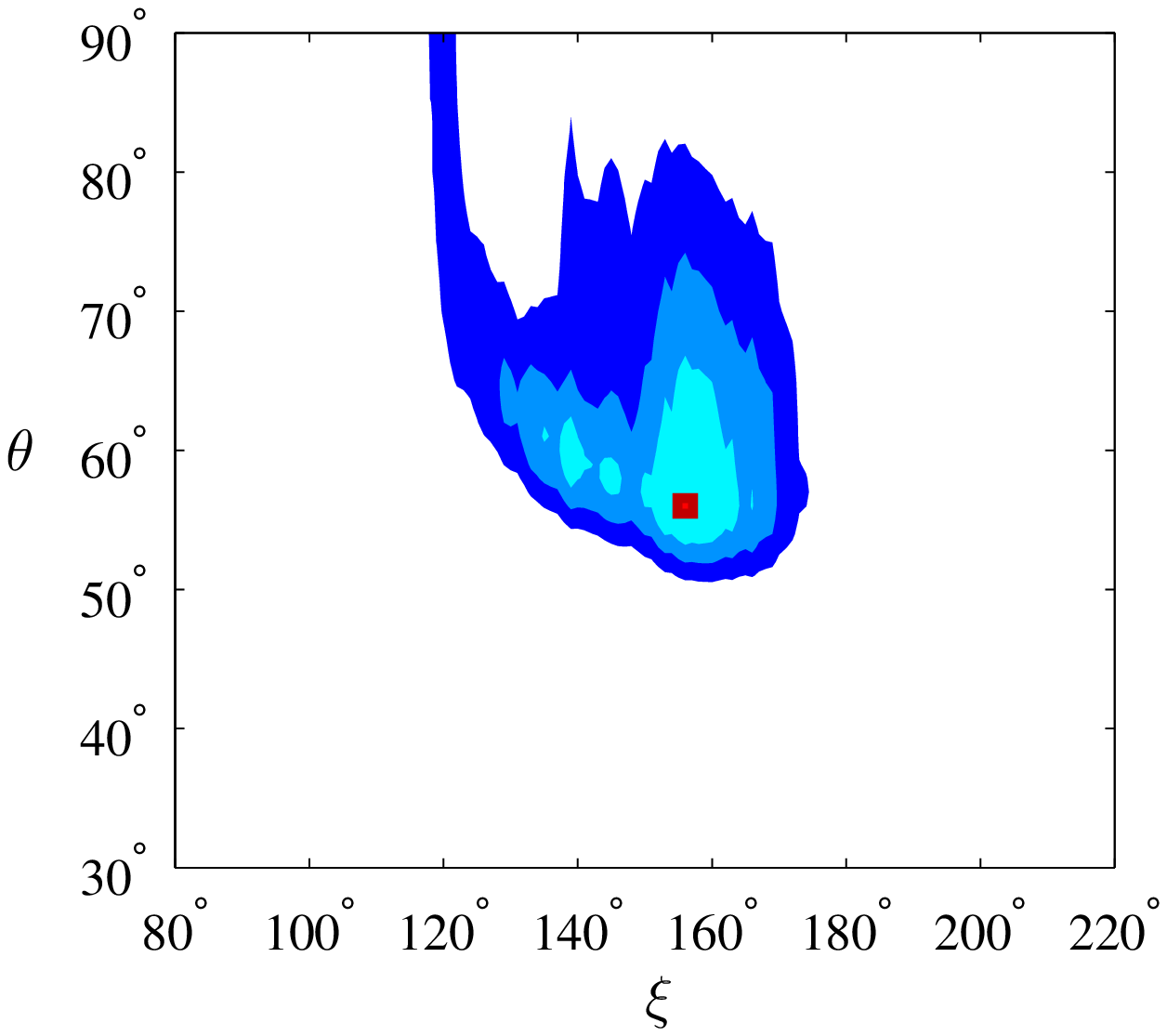,width=0.32\textwidth}
\epsfig{figure=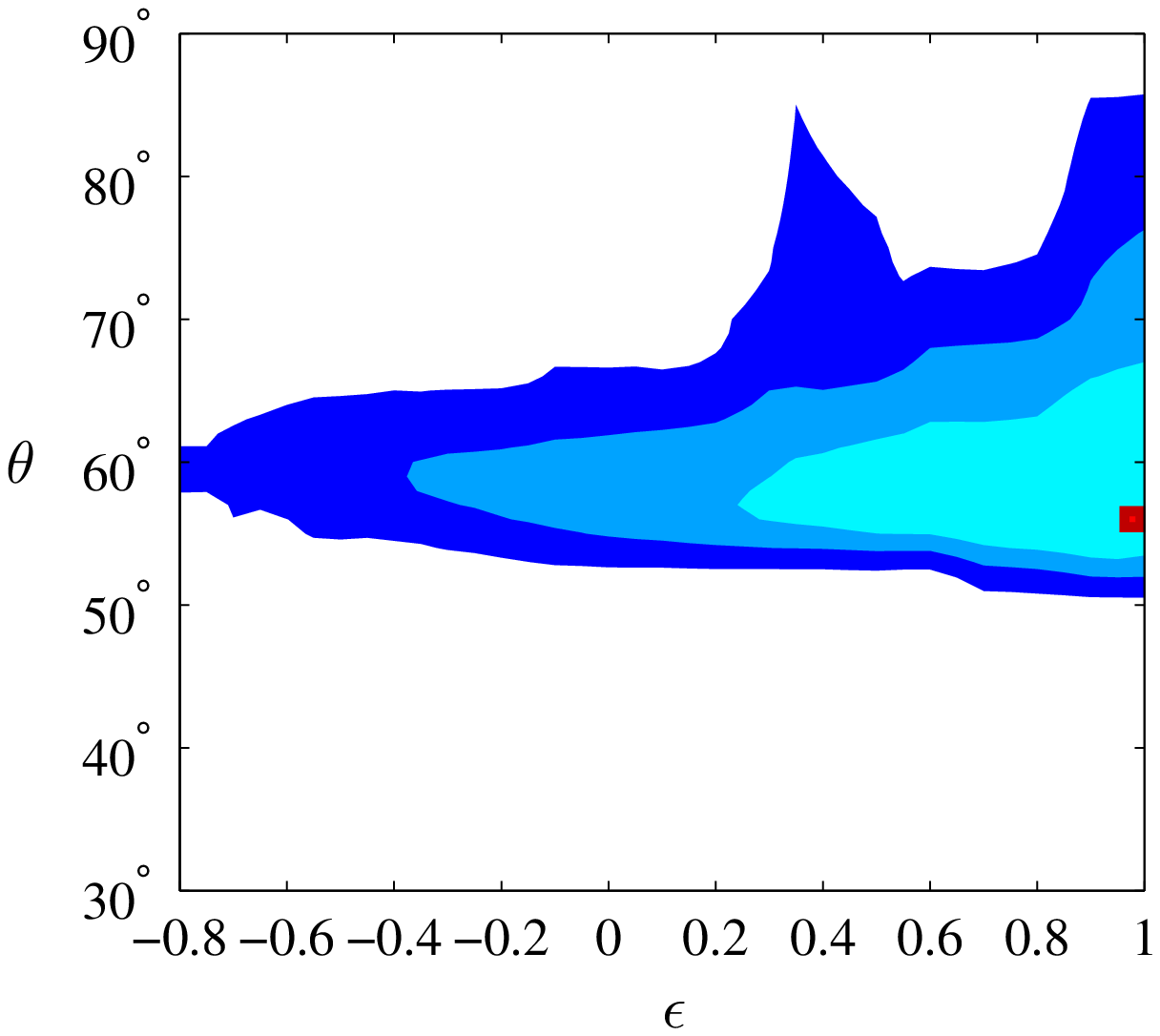,width=0.32\textwidth}
\epsfig{figure=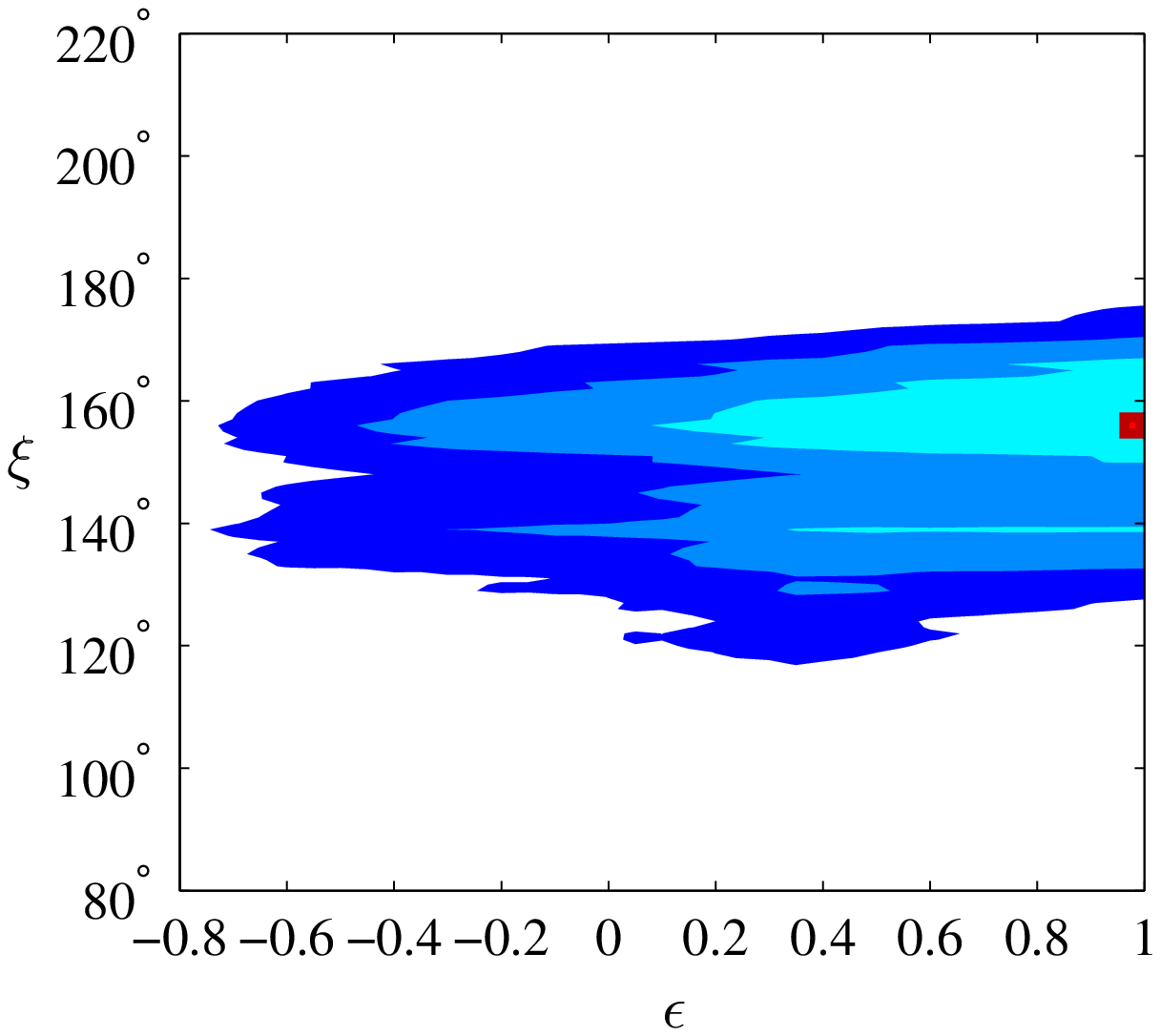,width=0.32\textwidth}
\epsfig{figure=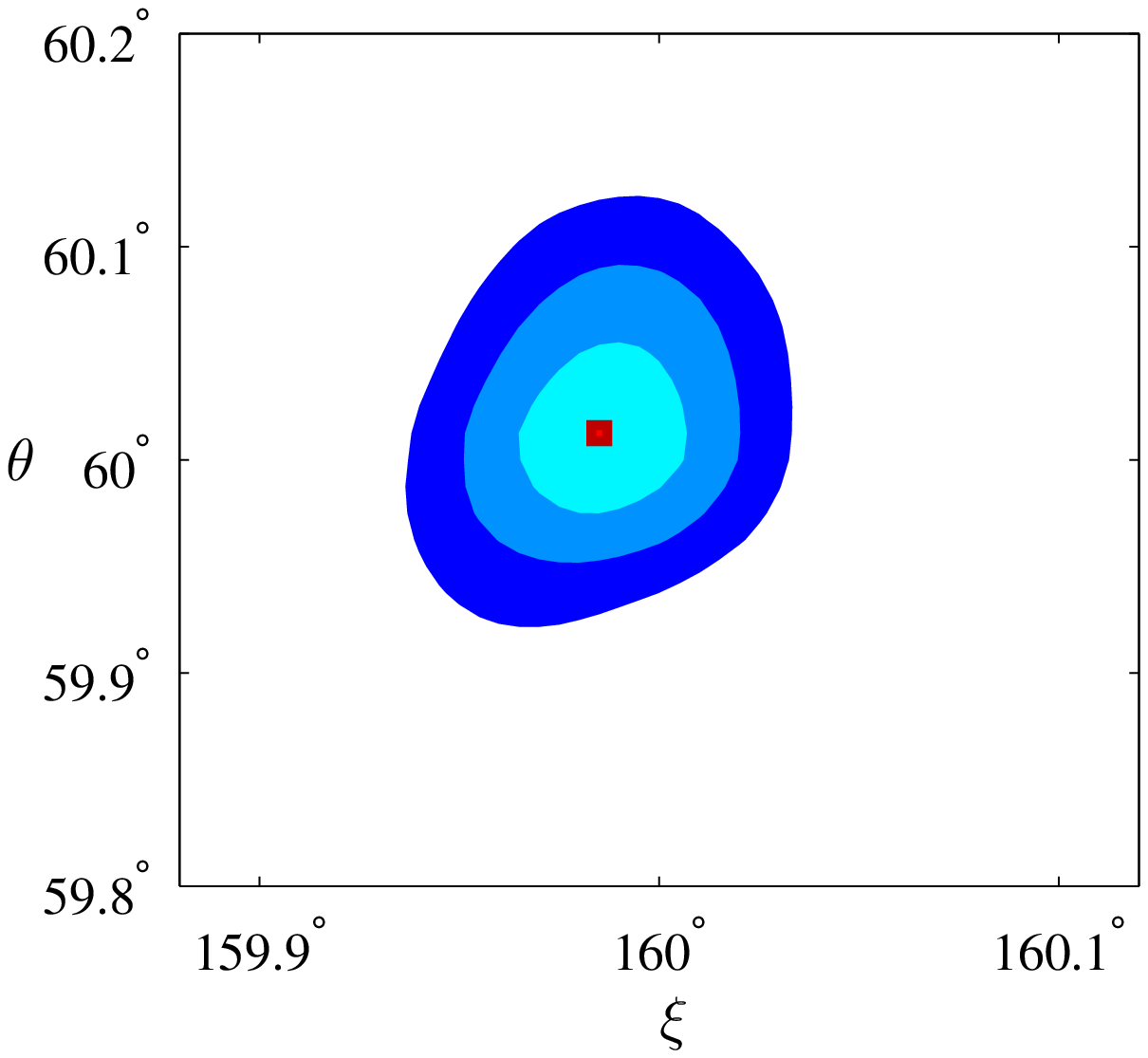,width=0.32\textwidth}
\epsfig{figure=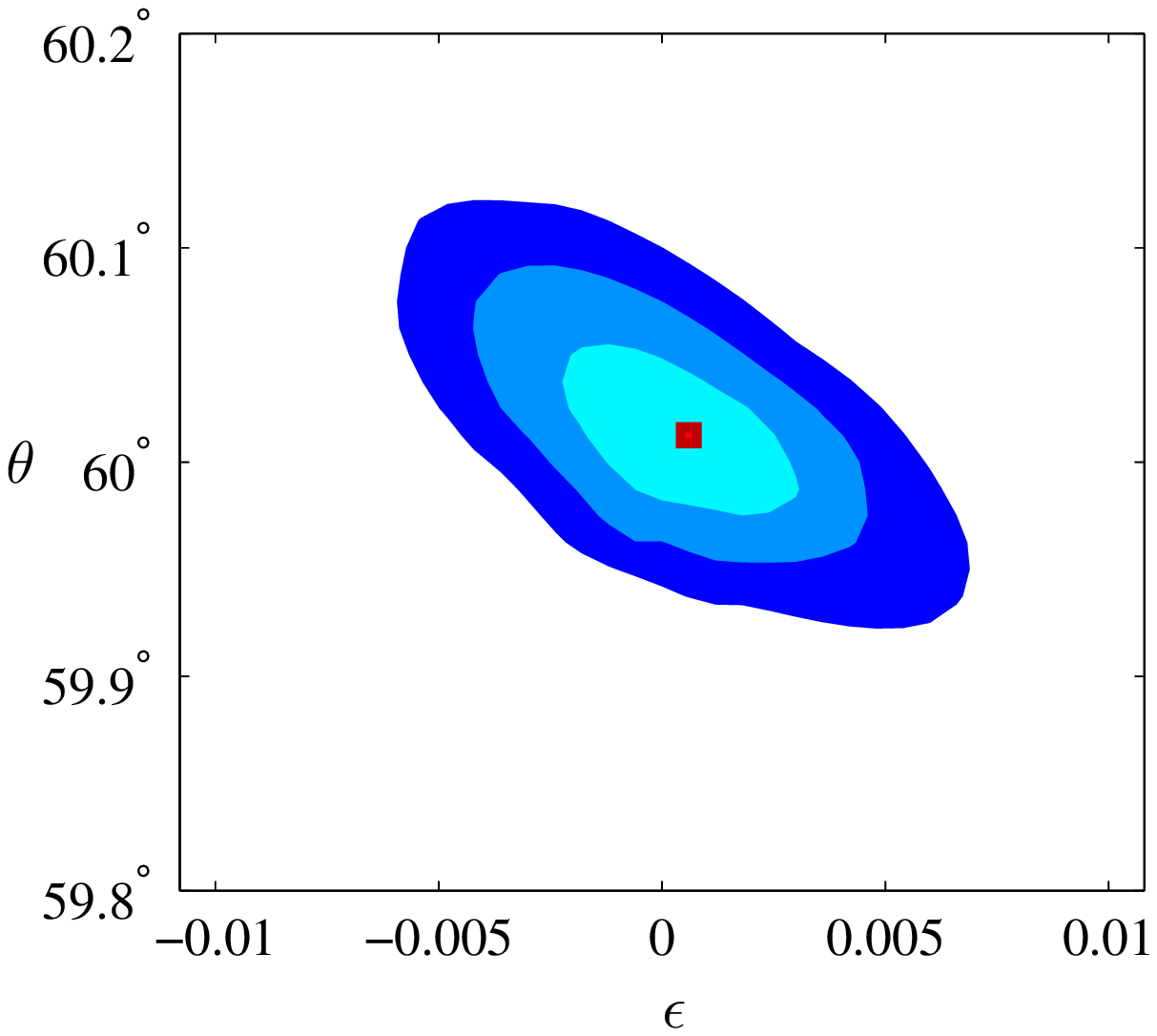,width=0.32\textwidth}
\epsfig{figure=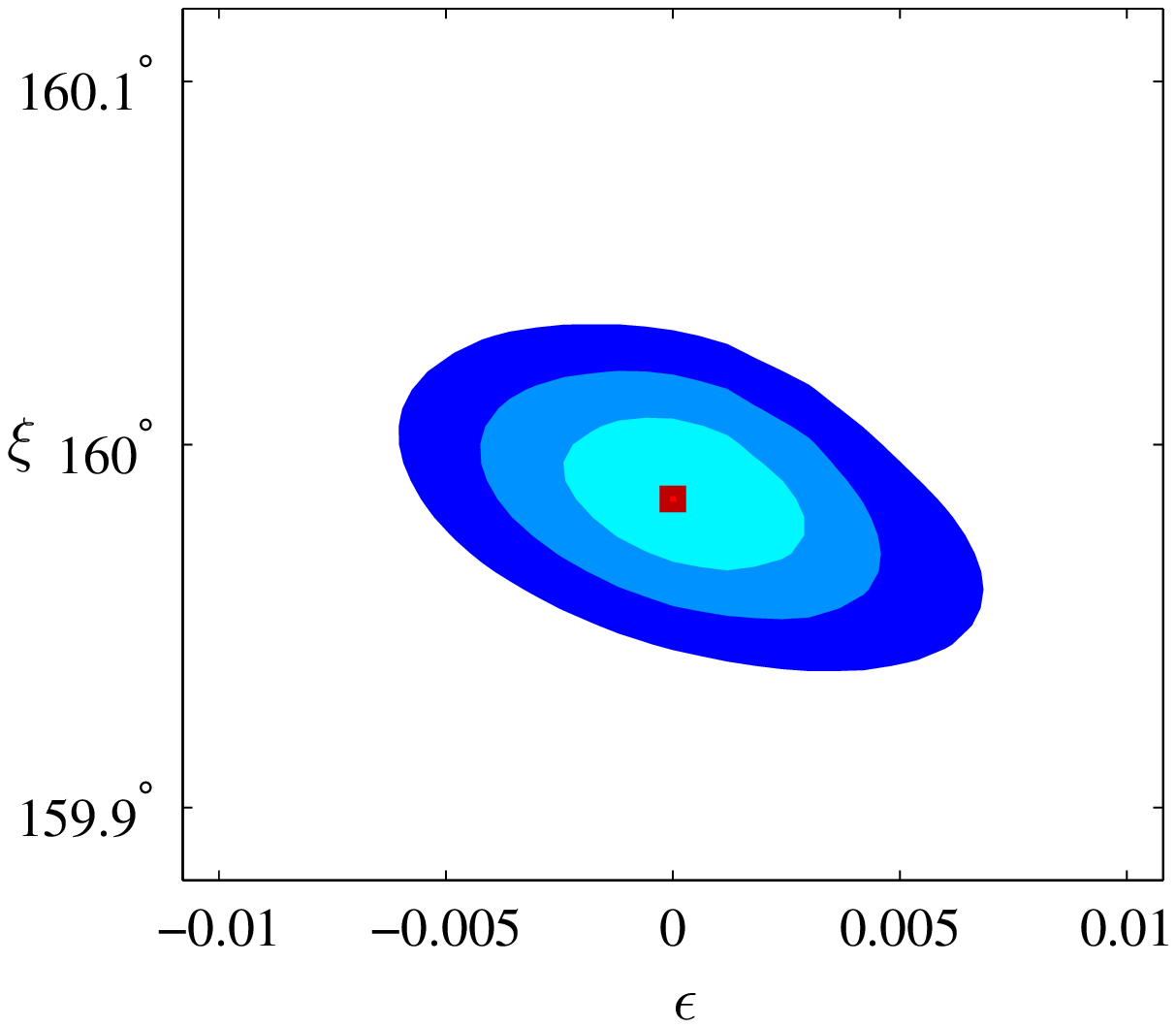,width=0.32\textwidth}
\end{center}
\caption{$1\sigma$, $2\sigma$, and $3\sigma$ confidence contours of the posterior probability density as a function of (left panels) inclination $\theta$ and the spin orientation $\xi$, (center panels) the inclination and the deviation parameter $\epsilon$, and (right panels) the spin orientation and the deviation parameter, marginalized over all other quantities. The top row panels correspond to the current constraints on these parameters from existing EHT data, while the bottom row panels correspond to our simulation of near-future EHT observations. The red dot in each panel denotes the maximum of the respective 2D probability density.}
\label{fig:constraints2}
\end{figure*}

{\it Context of other Kerr-like metrics ---} Several Kerr-like metrics have been proposed to date (e.g., \cite{GB06,metrics,Jmetric}). The quasi-Kerr metric derives from the Hartle-Thorne metric~\cite{HTmetric} which was originally developed for the description of neutron stars. The quasi-Kerr metric is a solution of the vacuum Einstein equations for spins that satisfy $|a_\ast| \ll 1$, provided $\epsilon$ is small. Here, however, we treat the quasi-Kerr metric as an ``exact'' metric as discussed in Ref.~\cite{pathologies} and study the impact of a deformed quadrupole moment on black hole accretion flows. Thus, we neither require the spin $a_\ast$ nor the deviation parameter $\epsilon$ to be small.

As shown in Ref.~\cite{pathologies}, the quasi-Kerr metric actually harbors a naked singularity as well as pathological regions of space around this singularity where closed timelike curves exist and Lorentzian symmetry is violated. Therefore, as in Ref.~\cite{Bro14}, we impose a cutoff radius at $r=3r_g$, which encloses all unphysical regions, and we consider all photons and matter particles that pass through this radius ``captured,'' i.e., they no longer contribute to our simulation. With this setup, the quasi-Kerr metric effectively describes a black hole. In addition, we only consider values of the spin and the parameter $\epsilon$ for which the ISCO lies at a radius $r\geq4r_g$. These restrictions define the excluded region shown as the gray region in the bottom panels of Figs.~\ref{fig:constraints} and \ref{fig:sim} and ensure that the quasi-Kerr part of the metric is always much smaller in magnitude than the Kerr part. Consequently, our simulation actually tends to underestimate the effect of the deviation from the Kerr metric, because its impact would be the strongest at small radii which we partially exclude.

The quasi-Kerr metric has the advantage of being of a particularly simple form shortening the computational time required for extensive parameter studies (though this is not a critical limitation). In addition, while other Kerr-like metrics such as the one of Ref.~\cite{Jmetric} are much broader in scope and have physical properties that are better suited for tests of the no-hair theorem (see the discussion in Ref.~\cite{Jmetric}), the use of the quasi-Kerr metric allows us to establish direct contact between our results and the analysis of early EHT data of Ref.~\cite{Bro14} and highlights the tremendous improvement in precision achievable with larger EHT arrays. On the other hand, the metric of Ref.~\cite{Jmetric} can more easily accommodate large deviations from the Kerr metric which are favored in our current analysis and can be mapped to known black-hole solutions in certain alternative theories of gravity (see Refs.~\cite{ShadowSize2,Review1}). See Ref.~\cite{Review1} for a review on Kerr-like metrics and tests of the no-hair theorem with electromagnetic observations of Sgr~A$^\ast$.

In a different interpretation, the free parameter $\epsilon$ can also be regarded as a measure of the underlying systematic uncertainties in the measurement (see the discussion in Ref.~\cite{pathologies}). Comparing the results of our analysis of the current EHT data with the results of Ref.~\cite{update} which are based on the same RIAF model but assume the Kerr metric instead (i.e., $\epsilon=0$), the additional degree of freedom in terms of the parameter $\epsilon$ points to the presence of substantial systematic uncertainties (linked to, e.g., variability, the morphology of the accretion disk, and potential outﬂows; see, also, the discussion in Ref.~\cite{update}) which need to be incorporated. While our analysis yields robust results for the inclination and spin orientation, the spin magnitude is unconstrained (cf., Refs.~\cite{EHTuncertainties1,EHTuncertainties2}).

{\it Parameter space ---} Our analysis of the current data considers the same parameter space as Ref.~\cite{Bro14} with upper bounds on the spin magnitude $a_\ast\leq0.9$ and deviation parameter $\epsilon\leq1$. These are conventions, chosen such that the quasi-Kerr part of the metric is always much smaller in magnitude compared to the Kerr part of the metric. A spin-dependent lower bound on the parameter $\epsilon$ is determined by the properties of the quasi-Kerr metric at small radii such that the ISCO is located at a radius $r\geq4r_g$ for any pair $(a_\ast,\epsilon)$. Since the RIAF model is probably not well defined for counter-rotating disks~\cite{flowmodels}, we also require $a_\ast\geq0$. For the inclination and the spin orientation, we allow for the ranges $30^\circ\leq\theta\leq90^\circ$ and $0^\circ\leq\xi\leq360^\circ$, respectively.

{\it Current and near-future constraints ---} Figure~\ref{fig:constraints2} shows the inclination--spin orientation, the inclination--quadrupolar deviation, and the spin orientation--quadrupolar deviation posterior probability distributions of our respective analyses of existing and simulated EHT data, each marginalized over all other parameters.


\begin{thebibliography}{10}
\bibitem{nocharge}
We expect astrophysical black holes to be electrically neutral since any net charge would neutralize quickly.
\bibitem{astar}
Here, $a_\ast\equiv a/M$ is the dimensionless spin parameter.
\end{thebibliography}
\end{document}